# LEADS: Lightweight Embedded Assisted Driving System

Tianhao Fu[1,2,4,*], Querobin Mascarenhas[1,2], Andrew Forti[1,2,3]

[1]Villanova College, King City, ON, Canada

[2]Project Neura, Toronto, ON, Canada

[3]University of Waterloo, Waterloo, ON, Canada

[4]Vector Institute, Toronto, ON, Canada

**Abstract.** With the rapid development of electric vehicles, formula races that face high school and university students have become more popular than ever as the threshold for design and manufacturing has been lowered. In many cases, we see teams inspired by or directly using toolkits and technologies inherited from standardized commercial vehicles. These architectures are usually overly complicated for amateur applications like the races. In order to improve the efficiency and simplify the development of instrumentation, control, and analysis systems, we propose LEADS (Lightweight Embedded Assisted Driving System), a dedicated solution for such scenarios.

Home: https://leads.projectneura.org

Docs: https://leads-docs.projectneura.org

Repository: https://github.com/ProjectNeura/LEADS

## 1 Introduction

LEADS is a lightweight embedded assisted driving system. It is designed to simplify the development of instrumentation, control, and analysis systems for racing cars. It is written in well-organized Python and C/C++ and has impressive performance. It is not only out-of-the-box but also fully customizable. It provides multiple abstract layers that allow users to pull out the components and rearrange them into a new project [1]. You can either configure the existing executable modules (LEADS VeC) simply through a JSON file or write your own codes based on the framework as easily as building a LEGO.

The hardware components chosen for this project are geared towards amateur developers. It uses neither a CAN bus (although it is supported) nor any dedicated circuit board, but generic development kits such as Raspberry Pi and Arduino instead. However, as it is a high-level system running on a host computer, the software framework has the ability to adapt to any type of hardware component with extra effort.

---

*Corresponding author: terry.fu@projectneura.org

## 2  Ecology

As stated in its name, LEADS tries to keep a modular structure with a small granularity to increase reliability [2] and so that users have more flexibility in dependencies and keep the system lightweight. To do so, we split LEADS into multiple packages according to the programming language, platform, features, and dependencies, but generally, these packages are grouped into 3 types: LEADS (LEADS framework), LEADS VeC, and accessories.

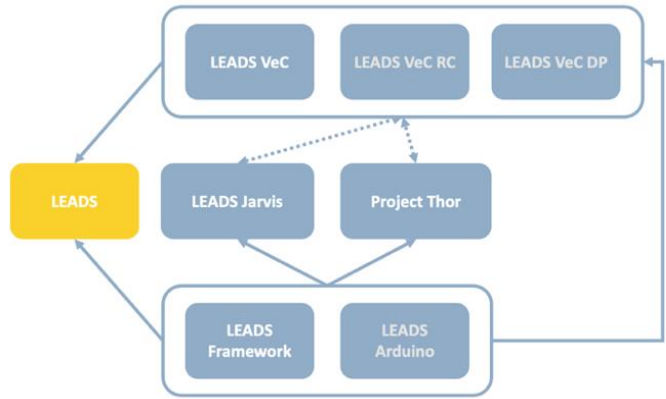

**Fig. 2.1.a.** LEADS Ecology

## 2.1  LEADS

In a narrow sense, LEADS refers to the LEADS framework, which consists of an abstract skeleton and implementations for various businesses. The framework includes a context that interacts with the physical car, a device tree system, a communication system, a GUI system, and many specifically defined devices.

## 2.2  LEADS VeC

LEADS VeC composes a set of executable modules that are designed for the VeC (Villanova Electric Car) Project. These modules are customizable and configurable to a certain extent, yet they have limitations due to some assumptions of use cases [1]. Compared to the LEADS framework, it sits closer to users and allows users to quickly test their ideas.

The following table displays its high performance by the maximum frame rate on various platforms when running LEADS VeC in emulation mode.

| Test Platform | CPU Cores | Clock Speed (GHz) | Maximum Frame Rate (FPS) |
|---|---|---|---|
| **Apple MacBook Pro (M3)** | 8 | 4.05 | 260 |
| **Orange Pi 5 Pro 8GB** | 8 | 2.4 / 1.8 | 200 |

| Raspberry Pi 5 8GB | 4 | 2.4 | 100 |
| --- | --- | --- | --- |
| Raspberry Pi 4 Model B 8 GB | 4 | 1.8 | 60 |

**Table. 2.2.a.** Maximum Frame Rate of LEADS VeC in Emulation Mode on Different Platforms

## 2.3  Accessories

Accessories contribute a huge portion to the richness of the LEADS ecosystem. These accessories include plugins that directly interact with the core programs, such as Project Thor, and standalone applications that communicate with LEADS through the data link, such as LEADS Jarvis.

# 3  Methods

## 3.1  Framework

The LEADS framework ensures that its applications, including LEADS VeC, have extremely high standards. They usually provide promising safety but still, always keep our Safety Instructions in mind.

Most of the codes are written in Python, and the dependencies are carefully chosen so that LEADS runs everywhere Python runs. In addition, on platforms like Arduino, where we must use other programming languages, we try hard to maintain consistency.

### 3.1.1 Device Tree

We organize all devices in a tree structure where each node can belong to only one parent node but may have multiple child nodes.

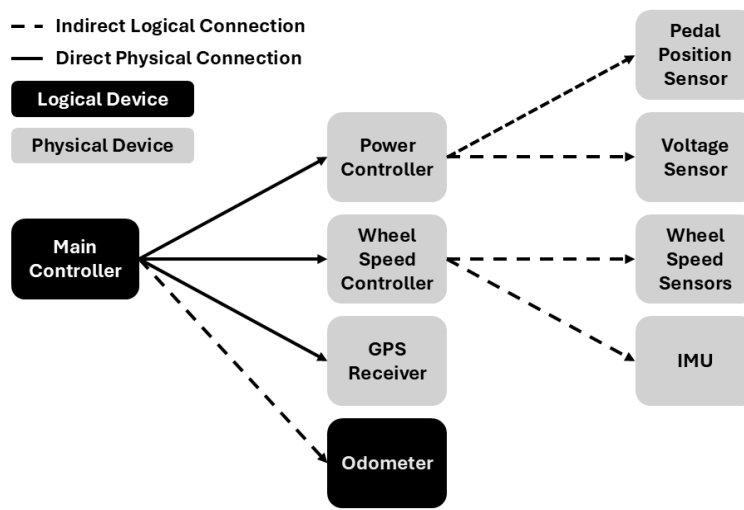

**Fig. 3.1.1.a.** Example Device Tree in LEADS VeC

In the global registry, all devices must have unique tags as identifiers. Controllers are a special variant of devices that can hold children. Each controller has a local registry to store the pointers to the child devices.

This tree mainly plays a role in dependency paths and error handling. In the previous example, the pedal position sensor relies on the existence of the power controller, and an error in the power controller is likely to result in errors in the pedal position sensor's readings.

## 3.1.2 Communication System

LEADS contains a communication system that focuses on business logic only and is independent of specific protocols. The communication system follows the C/S architecture as it is intended for and, by default, uses the TCP protocol [3]. The server end contains a listener thread (shown as "Thread 0" and can be the same as the main thread) and multiple processor threads that form a pool, compared to which the client is much simpler as it only has one connection thread (shown as "Thread 0" and can be the same as the main thread).

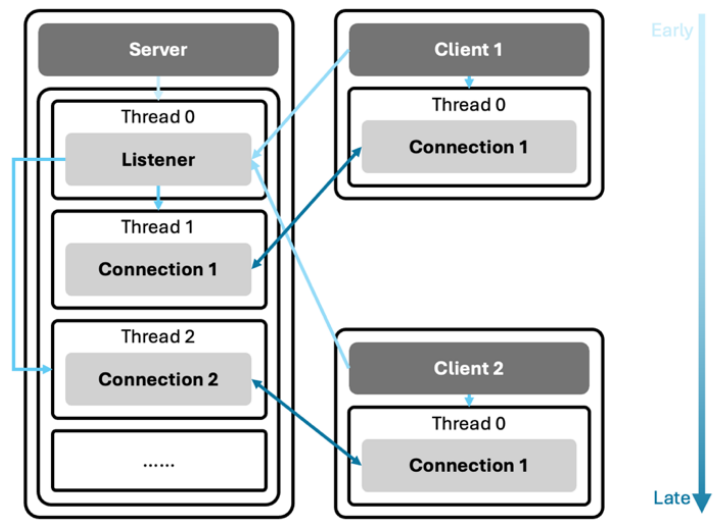

**Fig. 3.1.2.a.** Architecture of the Communication System

There are two classes in the communication system: entities and connections. Entities are services such as servers and clients that operate connections, whereas connections define the detailed procedure of operations. The philosophy of this system is that sending is the only active operation for both the server and the client. All other operations, including receiving, should be done passively through callback methods.

**Separator.** Long TCP connections do not split the streams into messages automatically. Hence, at the application level, we define a separator to do so. By default, LEADS uses ";" as the separator, yet in some scenarios where the message itself may contain the character, the separator can also be customized.

**Remainder.** Often network congestion causes multiple messages to be received at the same time. In order to return only one message to the interface at a time, the remaining messages must be stored in a buffer. This buffer is called "remainder".

**Serial Communication.** In addition to the default one, we also provide an implementation of connection for serial communication. It has the same APIs but requires a serial connection instead of a socket.

### 3.1.3 Data Persistence

In a lot of use cases, data needs to be stored in hard drives or cached in memory, yet due to the limited space and high cost of memory, often the data requires compression. As users ourselves, we deeply acknowledge the efforts to implement the process every time. Therefore, we integrate a toolkit for data persistence in LEADS. It works just like a regular list, but you can specify the maximum length and a compressor that is the mean compressor by default. The mean compressor reduces data memory usage by averaging adjacent numbers and merging them while ensuring the same integral.

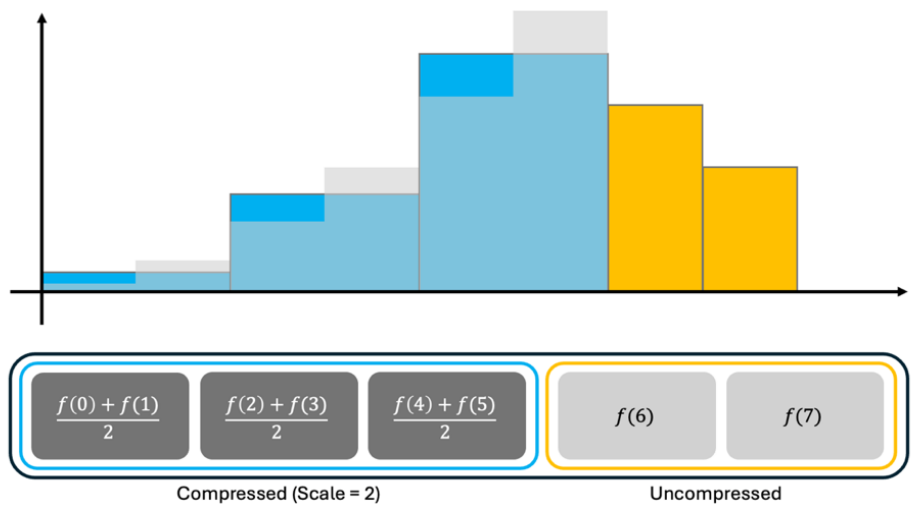

**Fig. 3.1.3.a.** Mean Compressor Algorithm

The integral is approximately given by the left Riemann sum $\int_b^a f(x)dx \approx \sum_{i=a}^b f(x_i) \cdot (x_{i+1} - x_i)$. With this algorithm, we can now cache, for example, speed data for a wide range of time in only a few kilobytes of memory and still be able to calculate the correct mileage.

The diagram below shows the relationship between the number of elements stored and the memory usage.

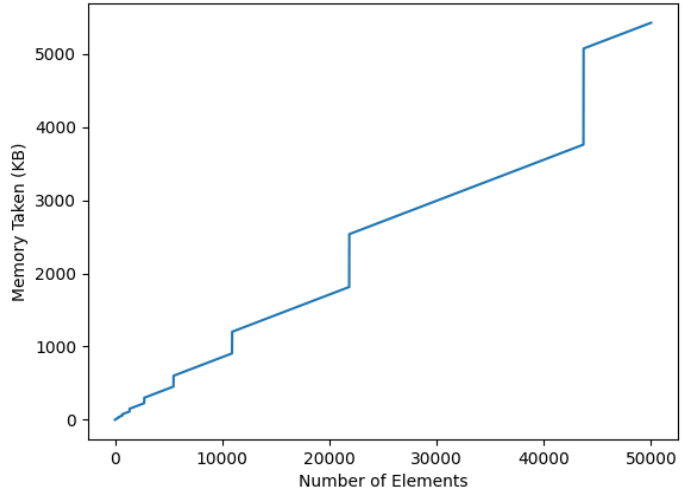

**Fig. 3.1.2.b.** Memory Taken over Number of Elements

## 3.2 Graphical User Interface

## 3.2.1 Performance Checker

LEADS GUI runs in the main thread at a constant refresh rate. However, just like with rendering in modern games, the additional latency introduced by frame operations means that the actual frame rate is often lower than expected. Therefore, we introduce the performance checker in LEADS GUI. It automatically adjusts the inter-frame interval by counting delays and predicting the next delay using a 5-degree polynomial regression. The performance checker stores both the delays and the net delays of the past 10 seconds, where the delay refers to the total time difference between two recorded frames, and the net delay refers to the duration of a frame. The inter-frame interval $w_t$ at time $t$ is given by $w_t = \frac{1}{r} - \max(\min(\text{E}(N|T)|_{x=t}, \frac{1}{r} - 0.001), 0)$ where $r$ is the target refresh rate, $N$ is the queue of net delays with $|N| = 10r$, and $T$ is the queue of times with $|T| = 10r$. All time-related variables are in seconds and $r$ is in FPS.

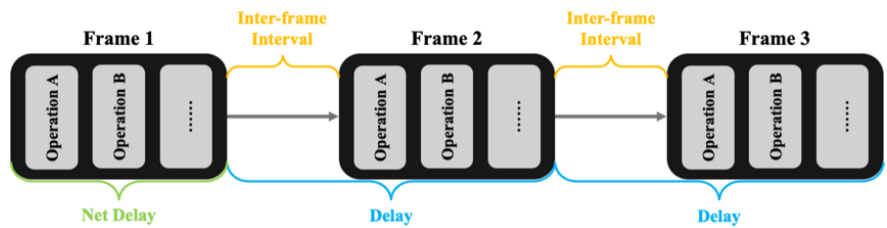

**Fig. 3.2.1.a.** Frame Structure

In our tests, on average, it takes about 30 seconds for the performance checker to stably maintain the target refresh rate. The experiment is conducted with LEADS VeC in emulation mode on an Apple MacBook Pro (M3). The tables and the graphs are concluded separately in 2 tests.

| Target Refresh Rate (FPS) | Frame Rate at 1s (FPS) | Frame Rate at 10s (FPS) | Frame Rate at 20s (FPS) | Frame Rate at 30s (FPS) |
|---|---|---|---|---|
| 30 | 26.32 | 28.83 | 29.88 | 30.00 |
| 60 | 48.11 | 56.83 | 59.79 | 60.01 |
| 120 | 75.52 | 106.73 | 116.42 | 120.66 |
| 240 | 105.83 | 186.51 | 216.24 | 242.31 |

**Table. 3.2.1.a.** Frame Rates at Different Times under Various Target Refresh Rates

| Target Refresh Rate (FPS) | Avg Net Delay at 1s (ms) | Avg Net Delay at 10s (ms) | Avg Net Delay at 20s (ms) | Avg Net Delay at 30s (ms) |
|---|---|---|---|---|
| 30 | 6.88 | 3.24 | 3.19 | 3.34 |
| 60 | 4.71 | 2.24 | 1.77 | 1.69 |
| 120 | 5.09 | 1.33 | 1.47 | 2.27 |
| 240 | 5.35 | 1.67 | 1.26 | 2.00 |

**Table. 3.2.1.b.** Average Net Delays at Different Times under Various Target Refresh Rates

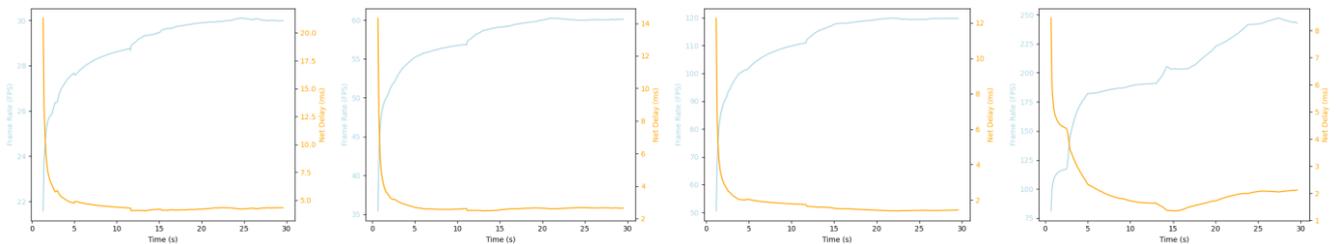

**Fig. 3.2.1.b.** Frame Rate and Net Delay over Time under Different Target Refresh Rates

## 3.2.2 Canvas-based Widgets

Most GUI libraries provide widgets that are only designed for single purposes. For example, typography only displays text. All meters require a canvas to display the rich content. When switching between different meters, destroying and recreating these widgets allocate excess resources. Therefore, we abstract each instrument into a set of rendering methods: a raw renderer and a dynamic renderer. These methods can be applied to any canvas. The widget object itself only stores style-related arguments. The raw renderer draws static elements, such as the background, whereas the dynamic renderer draws updated elements, such as speed. All instruments in the same position, which are interchangeable, share one single canvas instance.

## 3.3 ESC Systems

Plugins can be easily customized and installed in LEADS. It also comes with several existing ones, including four ESC plugins that realize four ESC systems. All four systems have four calibrations: standard, aggressive, sport, and off. Each intervention comes later than the previous one, respectively.

### 3.3.1 DTCS (Dynamic Traction Control System)

DTCS helps you control the amount of rear wheel slip by detecting and comparing the speed difference between all wheels. It allows a certain amount of drift while ensuring grip.

It reduces the throttle output if the speed of the driven wheels is significantly lower than the speed of the driving wheels.

**AWD (All-wheel Drive) Vehicles.** In AWD vehicles, the determination of slip gets complicated as there is no distinction between driven and driving wheels. Most AWD vehicles are equipped with mechanical traction control systems instead of software implementations [4]. Without any differential device, our solution is to use GPS to identify the real velocity, whereas the rear-wheel-drive solution relies on the driven wheels.

### 3.3.2 ABS (Anti-lock Braking System)

ABS allows the driver to safely step on the brakes to the bottom without locking the brakes. Although this may increase braking distances, it ensures you always have the grip.

Simply put, if the front wheel speed drops sharply to close to 0, it reduces the braking force.

### 3.3.3 EBI (Emergency Braking Intervention)

EBI is available on most modern family cars. It actively applies the brakes at the limit of the braking distance. This system greatly reduces the probability of rear-end collisions.

### 3.3.4 ATBS (Automatic Trail Braking System)

ATBS monitors the steering angle and adjusts the brakes in time to ensure that the front of the car obtains the corresponding downforce. Its intervention will avoid under-steers (pushing the head) or over-steers (drifting).

The downforce on each wheel, assuming a four-wheel model, can be approximated by $CFC = \frac{mg}{4} \pm \frac{mha_{forward}}{2w} \mp \frac{mha_{lateral}}{2L}$ where $m$ is the mass of the car, $h$ is the height of the center of mass of the car, $a_{forward}$ is the forward acceleration (positive forward and negative backward), $a_{lateral}$ is the lateral acceleration (positive rightward, negative leftward), $w$ is the width of the car, and $L$ is the length of the car.

Specifically, $CFC_{front\ left} = \frac{mg}{4} - \frac{mha_{forward}}{2w} + \frac{mha_{lateral}}{2L}$, $CFC_{front\ right} = \frac{mg}{4} - \frac{mha_{forward}}{2w} - \frac{mha_{lateral}}{2L}$, $CFC_{rear\ left} = \frac{mg}{4} + \frac{mha_{forward}}{2w} + \frac{mha_{lateral}}{2L}$, and $CFC_{rear\ right} = \frac{mg}{4} + \frac{mha_{forward}}{2w} - \frac{mha_{lateral}}{2L}$.

## 3.4 Sensors

### 3.4.1 Wheel Speed Sensor

In any modern wheel-based vehicle, wheel speed sensors are of great importance for determining the vehicle's speed, and most ESC systems require an accurate speed of each wheel [5]. The most classic type of wheel speed sensor relies on the Hall effect [6]. The sensor is effectively a Hall switch that outputs a pulse when a magnet is detected within a certain range. By measuring the time interval between two pulses triggered by one or more magnets mounted on the wheel or equivalent spinning mechanism, we can calculate the frequency and therefore determine the speed according to $v = \frac{60c}{t_r n}$ where $v$ is the speed in kilometers per hour, $c$ is the wheel circumference in centimeters, $t_r$ is the time interval between two pulses, or the time of one revolution, in seconds, and $n$ is the number of magnets (assuming they are equally distributed). However, as it is shown in the algorithm, the rate of return is proportional to the speed at which the wheel is rotating. This defect can be eased by methods such as frequency-domain adaptive filtering [7].

**Pulse Normalization.** Each magnet may be within the detection range for more than one tick in one revolution. Therefore, we only capture the instant change in voltage from high (off) to low (on), which is the left end of a square wave.

**False Pulses Dropout by Acceleration.** Another limitation of the design is that the switch may be off and on again shortly due to bumping. We proposed a correction method that drops out outlier pulses by comparing acceleration. The predicted speed $v_p$ in $m/s$ at time $t_1$ is given by $v_p(t_1) = v(t_0) + \int_{t_0}^{t_1} a(t)dt \approx v(t_0) + 1.8(a(t_0) + a(t_1))(t_1 - t_0)$ where $a(t)$ is acceleration measured in $m/s^2$, and $t$ is time measured in seconds. If $\left|\frac{v(t_1) - v_p(t_1)}{v_p(t_1) - v(t_0)}\right| > \lambda$ where $\lambda$ is the threshold factor that is 1.5 by default, then drop the frame at $t_1$. This is not a disturbance to the ESC systems because the wheel speeds only provide a threshold that binarily triggers those systems and the factor $\lambda$ leaves enough room for those systems to be activated.

### 3.4.2 Accelerometer

Accelerometers, sometimes referred to also as G force meters, are often seen on sports cars. As an example, the Adafruit BNO085 we use is a 9-DOF orientation fusion IMU (Inertial Measuring Unit) that provides promising accuracy [8]. However, because IMU is based on the inertial system, it is disturbed by the normal force against gravity at rest. When the unit is not perfectly horizontal, which is most likely the case, the component of that normal force in the sensor's reference frame will cause the acceleration it measures to be no longer equal to the derivative of velocity. Therefore, we need to eliminate that disturbance from the sensor's output to get linear acceleration. To do this, we apply transformations in a Euclid space to make gravity in the sensor's relative using rotation component matrices that are defined as $R_x(\phi) =$

$\begin{bmatrix} 1 & 0 & 0 \\ 0 & \cos\phi & -\sin\phi \\ 0 & \sin\phi & \cos\phi \end{bmatrix}$, $R_y(\theta) = \begin{bmatrix} \cos\theta & 0 & \sin\theta \\ 0 & 1 & 0 \\ -\sin\theta & 0 & \cos\theta \end{bmatrix}$, and $R_z(\psi) = \begin{bmatrix} \cos\psi & -\sin\psi & 0 \\ \sin\psi & \cos\psi & 0 \\ 0 & 0 & 1 \end{bmatrix}$ where $\psi$, $\theta$, and

$\phi$ are yaw, pitch, and roll, respectively, such that the rotation matrix $R = R_z(\psi) \cdot R_y(\theta) \cdot R_x(\phi)$. The gravity in the sensor's relative is then $\overrightarrow{F_g} = R^T \cdot \vec{g}$ where $R^T$ is the transpose of $R$.

## 3.5 Hardware Schema

Normally, modern vehicles use CAN buses across one central ECU and multiple nodes. However, for our specific targeted use cases, we recommend GPIO SBCs enhanced by MCUs. Almost every GPIO SBC contains an Arm64 SoC that can run Linux. This greatly lowers the barrier to entry for creating high-resolution user interfaces, which is a necessity in modern aesthetics. However, the multithreading model in Linux introduces intervals where a thread is frozen [9]. If active sensors like wheel speed sensors happen to be triggered within that interval, the program cannot capture the pulse. In addition, the lack of DAC (Digital-to-analog Converter) is another problem. Therefore, using serial communication over USB, MCUs such as Arduinos are connected to the main computer for these tasks.

## 3.6 Data Analysis

### 3.6.1 Inferences

Inferences aim to supplement the missing dimensions in the original data by inferring other dimensions. Each inference iterates the whole dataset while caching a certain number of frames and skipping frames that do not require supplementary. Since some may allocate massive amounts of computing time and memory, it is not recommended to use them in real time.

**Safe Speed Inference.** Infer the speed based on the front wheel speed or the rear wheel speed by $v = \min(fws, rws)$.

**Speed Inference by Acceleration.** Infer the speed based on the acceleration by $v = \int a(t)\mathrm{d}t$.

**Speed Inference by Mileage.** Infer the speed based on the mileage by $v = \frac{\mathrm{d}s}{\mathrm{d}t}$.

**Speed Inference by GPS Ground Speed.** Infer the speed based on the GPS ground speed.

**Speed Inference by GPS Position.** Infer the speed based on the GPS position. This is equivalent to inferring the mileage based on GPS position and then inferring the speed based on the mileage.

**Forward Acceleration Inference by Speed.** Infer the forward acceleration based on the speed by $\vec{a} = \frac{\mathrm{d}v}{\mathrm{d}t}$. Note that this is not always reliable because speed is a scalar, but forward acceleration is not. Accelerating in reverse will still be counted as forward acceleration.

**Mileage Inference by Speed.** Infer the mileage based on the speed by $s = \int v(t)\mathrm{d}t$.

**Mileage Inference by GPS Position.** Infer the mileage based on the speed by $s = s_0 + \Delta s$.

**Visual Data Realignment by Latency.** Offset the delay introduced by camera recording and video encoding so that the sensor data and the picture of the exact frame match.

### 3.6.2 Statistics

**Trip-level Analysis.** LEADS iterates the whole race to bake some global variables such as absolute minimums and maximum values and, most importantly, the map.

**Lap-level Analysis.** After the trip data is baked, we split it into laps based on geometric location. The basic idea is to set up a collision volume and detect collisions between the vehicle's position and the path it passes through.

## Contributions

- **Tianhao Fu**: First Author, Project Initiative, Code Implementation
- **Querobin Mascarenhas**: Project Initiative, Team Management, Budget Management
- **Andrew Forti**: Team Management

## Funding

## Conflict of Interest

The authors declare no conflict of interest.